\documentclass[aps,twocolumn,superscriptaddress,showpacs,showkeys]{revtex4-2}
\usepackage{amsmath}
\usepackage{color,xcolor}
\usepackage{graphicx}

\newcommand{\bastar}{\begin{eqnarray*}}
\newcommand{\eastar}{\end{eqnarray*}}
\newskip\humongous \humongous=0pt plus 1000pt minus 1000pt

\newif\ifdtup

\relax
\newcommand{\W}{{\vec W}}
\newcommand{\n}{\hat n}

\newcommand{\hD}{{\hat D}}
\newcommand{\bD}{{\bar D}}
\newcommand{\bea}{\begin{eqnarray}}
\newcommand{\eea}{\end{eqnarray}}
\newcommand{\pd}{\partial}

\newcommand{\hB}{{\hat B}}

\newcommand{\G}{{\vec G}}
\newcommand{\B}{{\vec B}}

\newcommand{\vphi}{{\vec \phi}}
\newcommand{\vPhi}{{\vec \Phi}}

\newcommand{\hG}{{\hat G}}
\newcommand{\bA}{{\bar A}}
\newcommand{\bF}{\bar F}

\newcommand{\be}{\bar e}
\newcommand{\mn}{{\mu\nu}}
\newcommand{\om}{\omega}

\newcommand{\lam}{\lambda}

\newcommand{\vsig}{{\vec \sigma}}
\newcommand{\nn}{\nonumber}

\newcommand{\cD}{{\cal D}}

\newcommand{\cL}{{\cal L}}

\begin{document}
\title{Non-Abelian Magnon Gauge Interactions in Condensed Matter Physics}

\author{Y. M. Cho}
\email{ymcho0416@gmail.com}
\affiliation{Center for Quantum Spacetime, 
Sogang University, Seoul 04107, Korea}  
\affiliation{School of Physics and Astronomy,
Seoul National University, Seoul 08826, Korea}
\author{Franklin H. Cho}
\email{cho.franklin@qns.science}
\affiliation{Center for Quantum Nano Science,
Ewha Woman's University, Seoul 00000, Korea}

\begin{abstract}
We discuss three different but closely related theories which could describe varieties of condensed matters, 
in particular the frustrated magnetic materials and 
the multi-gap (ferro)magnetic superconductors with 
or without the photon-magnon mixing, where the genuine non-Abelian magnon gauge interaction plays the central role. The charactristic features of these theories 
are the existence of long range magnetic order and 
the spin-spin interaction described by the exchange of the messenger bosons, not by the instantaneous action at a distance. These theories could play important roles in our understanding of non-Abelian condensed matters and make the non-Abelian gauge interaction a main stream 
in the low energy physics. We discuss the physical implications of our results.     
\end{abstract}


\maketitle

{\bf Introduction}---It is well known that all four fundamental interactions in nature are described by gauge theories, Abelian and non-Abelian. In fact, except the electromagnetic interaction, they are described by non-Abelian gauge interactions. This tells that the non-Abelian gauge interaction is the fundamental interaction in nature. On the other hand, this non-Anbelian gauge interaction has rarely been used in condensed matter physics so far. This appears very strange. Of course the non-Abelian structures and the resulting non-Abelian topological objects have become common in condensed matter physics these days, and there have been huge amount of literatures on non-Abelian condensed matters, in particular 
on non-Abelian superconductivity and non-Abelian topological objects in condensed matters \cite{wen,zhong,prb05,prb06,epjb08}. Nevertheless 
the genuine non-Abelian gauge interaction has been almost non-existent in condensed matters. This is puzzling. 

Recently, however, there have been two 
remarkable new developments on the non-Abelian gauge interaction in condensed matters. First, 
a non-Abelian two-gap ferromagnetic suprerconductivity made of the spin-up and spin-down Cooper pairs described by an SU(2)xU(1) gauge theory which can be viewed as a non-Abelian generalization of the well-known ordinary Landau-Ginzburg theory of superconductivity 
has been proposed \cite{ap24,pla23}. Second, 
a non-Abelian SU(2) gauge theory described by 
the non-Abelian magnons in the frustrated magnetic materials has been proposed by Zarzuela and 
Kim \cite{zarkim}. The unique feature of these 
works is that for the first time they are advocating the existence of the genuine real non-Abelian gauge interaction in condensed matter physics.    
 
{\it The purpose of this Letter is to show that 
these two developments are not separate incidents 
but the same development which strongly indicates 
the possible existence of the genuine non-Abelian 
gauge interaction in condensed matter physics which could describe varieties of condensed matters including multi-gap ferromagnetic superconductors, frustrated magnetic materials, and spin liquids. To support this proposition, 
we present three possible non-Abelian gauge interactions in condensed matters, frustrated magnetic materials, two-gap ferromagnetic superconductors, and three-gap magnetic superconductors. We show that in all three 
cases the underlying physics is described by 
the non-Abelian magnon gauge interaction.} 

This strongly implies that genuine non-Abelian gauge interactions could indeed play important roles in condensed matter physics. In the following we discuss three possible non-Abelian gauge interactions in condensed matters.

{\bf Non-Abelian magnon interaction in frustrated magnetic material}---Consider a frustrated magnetic material made of the spin triplet 
magnets described by a real SU(2) triplet $\vphi=(\phi_1,\phi_2,\phi_3)$ interacting with the non-Abelian magnon. The strong magnetic interaction could be described by the following Lagrangian
\begin{gather}
\cL_1 =-\frac12 |D_\mu \phi|^2 -V(\vphi)
-\frac14 \G_\mn^2,    \nn \\
D_\mu \vphi=(\pd_\mu +g' \B_\mu \times \big) \vphi, 
~~V(\vphi)=\frac{\lambda}{2}\big(|\phi|^2
-\frac{\mu^2}{\lambda}\big)^2,  
\label{lag1}
\end{gather}
where $V(\vphi)$ is the self interaction potential of the spin triplet magnet which we choose to be quartic for simplicity, and $g'$, $\B_\mu$, $\G_\mn$ are the coupling constant and non-Abelian magnon gauge fields. Clearly the non-Abelian magnon interaction described by the Lagrangian is the type of gauge interaction proposed by Zarzuela and Kim in the frustrated magnetic 
materials \cite{zarkim}.

To simplify the Lagrangian we let $(\n_1,\n_2,\n_3)$ to be an arbitrary orthonormal SU(2) basis, and express $\vphi$ by 
\begin{gather}
\vphi= \rho~\n,~~~(\rho=|\vphi|,~~~\n^2=1).
\end{gather}
Now, selecting $\n$ to be the Abelian direction 
we make the Abelian decomposition of $\B_\mu$ to 
the restricted part $\hB_\mu$ and the valence 
part $\W_\mu$ \cite{prd80,prl81},
\begin{gather}
\B_\mu =\hB_\mu +\W_\mu,   \nn\\
\hB_\mu = B_\mu~\n -\frac1g \n \times \pd_\mu \n,
~~~B_\mu= \B_\mu \cdot \n,   \nn\\
\W_\mu=W_\mu^1 \n_1+W_\mu^2 \n_2.
\label{cdec}
\end{gather} 
where $\hB_\mu$ is the restricted magnon potential which satisfies the condition $D_\mu \n=0$, 
the potential which makes the Abelian direction $\n$ covariantly constant. Notice that $\hB_\mu$ has a dual structure, made of the non-topological Maxwellian part $B_\mu \n$ and the topological Diracian part $-(1/g) \n\times \pd_\mu \n$. 
The advantage of the Abelian decomposition is 
that it is gauge independent. Once the Abelian direction is selected the decomposition follows automatically, independent of gauge. Moreover, 
the restricted potential $\hB_\mu$ inherits 
the full non-Abelian gauge degrees of freedom inspite of the fact that it is restricted,
while the valence potential $\W_\mu$ becomes 
gauge covariant \cite{prd80,prl81}. 

From the Abelian decomposition (\ref{cdec}) 
we have
\begin{gather}
\G_\mn=\hG_\mn + \hD _\mu \W_\nu 
- \hD_\nu \W_\mu + g' \W_\mu \times \W_\nu,   \nn\\
\hD_\mu=\pd_\mu+g' \hB_\mu \times,   \nn\\
\hG_\mn= \pd_\mu \hB_\nu-\pd_\nu \hB_\mu
+ g \hB_\mu \times \hB_\nu =G_\mn' \n, \nn \\
G'_\mn=G_\mn + H_\mn
= \pd_\mu B_\nu'-\pd_\nu B_\mu',  \nn\\
G_\mn =\pd_\mu B_\nu-\pd_\nu B_\mu, \nn\\
H_\mn =-\frac1{g'} \n \cdot (\pd_\mu \n 
\times\pd_\nu \n) =\pd_\mu C_\nu-\pd_\nu C_\mu,  \nn\\
B_\mu' = B_\mu+ C_\mu,
~~~C_\mu =-\frac1{g'} \n_1\cdot \pd_\mu \n_2.
\end{gather}
Notice that the restricted field strength $\hG_\mn$ inherits the dual structure of $\hB_\mu$, so that it can also be described by 
two Abelian potentials, the Maxwellian $B_\mu$ 
and the Diracian $C_\mu$. But here the potential $C_\mu$ for $H_\mn$ is determined uniquely up 
to the $U(1)$ gauge freedom which leaves $\n$ invariant. This is the Abelian decomposition 
of the SU(2) gauge field known as the Cho decomposition, Cho-Duan-Ge (CDG) decomposition, 
or Cho-Faddeev-Niemi (CFN) 
decomposition \cite{fadd,shab,kondo}. 

With this we have
\begin{gather}
D_\mu \vphi =(\pd_\mu \rho)~\n 
+g' \rho~\W_\mu \times \n,
\end{gather}
so that the Lagrangian (\ref{lag1}) is expressed by
\begin{gather}
\cL_1 =-\frac 12 \pd_\mu \rho^2 -\frac{\lambda}{2}\big(\rho^2 -\rho_0^2 \big)^2
-\frac14 {G'}_\mn^2    \nn\\
-\frac14 (\hD_\mu\W_\nu-\hD_\nu\W_\mu)^2 
-\frac{g^2}{2} \rho^2 \W_\mu^2 \nn\\
-\frac{g}{2} G'_\mn \n \cdot (\W_\mu \times \W_\nu)
-\frac{g^2}{4} (\W_\mu \times \W_\nu)^2,  
\label{lag2}
\end{gather}
where $\rho_0=\sqrt{2\mu^2/\lambda}$ is 
the magnitude of the vacuum expectation value of the Higgs scalar field. Notice that this Lagrangian is mathematically identical to (\ref{lag1}), so that it retains the full SU(2) gauge symmetry of (\ref{lag1}). This shows that two of the three massless off-diagonal magnons $\W_\mu$ acquire the mass by the Higgs scalar. This, of course, is the mass generation by the Higgs mechanism. But notice that the Abelian magnon $B'_\mu$ remains massless, which tells 
that the magnon gauge interaction described by (\ref{lag1}) has a long range magnetic interaction. 

The non-Abelian structure of the Lagrangian 
(\ref{lag1}) assure the existence of 
the non-Abelian topological objects in frustrated magnetic materials, as emphasized by Zarzuela 
and Kim \cite{zarkim}. Indeed, it is well known that the Lagrangian admits the non-Abelian magnetic vortex and the non-Abelian monopole 
made of the magnon.This must be clear because 
the Abelian direction $\n$ can describe 
the $\pi_1 (S^1)$ magnetic vortex topology 
and the $\pi_2(S^2)$ monopole topology. And this strongly implies the existence of a new non-Abelian Meissner effect generated by 
the magnon in frustrated magnetic materials.

At this point one may ask two questions. First,
should the frustrated magnetic materials have 
a long range magnon force? In the model described by the Lagrangian(\ref{lag1}), the answer is yes. 
But this may not always be the case, because we 
could make the the massless magnon massive modyfying the Lagrangian slightly without much difficulty. The more relevant question is related to the mass generation of the off-diagonal 
magnons by the Higgs mechanism. The popular Higgs mechanism tells that the mass generation takes place by the spontaneous symmetry breaking of 
the Higgs field. On the other hand, the magnetic interaction in frustrated magnetic materials are so strong that (it is generally thought that) even at zero temperature the magnetic field may not be aligned. If so how can the vacuum of the Higgs 
triplet break the symmetry spontaneously in frustrated magnetic materials to generate the spontaneous symmetry breaking?    

The answer is that the mass generation in the Higgs mechanism in truth does NOT require any symmetry breaking, spontaneous or not. This must be clear from the Lagrangian (\ref{lag2}), where the mass of $\W_\mu$ magnons comes from the vacuum value of $\rho$. Since this $\rho$ is a scalar field, it cannot break any symmetry. And 
the vacuum fluctuation of the magnetic field 
in the frustrated magnetic materials at zero temperature is given by the fluctuation of 
the orientation of $\n$, not by $\rho$. This 
shows that the mass generation by the spontaneous symmetry breaking is really a misleading interpretation of the Higgs mechanism \cite{ap24,pla23}.    
 
{\bf Non-Abelian magnon interaction in two-gap ferromagnetic superconductor}---Now, we show that 
the same non-Abelian magnon interaction could 
play a crucial role in two-gap ferromagnetic 
superconductors. To do this consider the following non-Abelian Landau-Ginzburg theory of two-gap SU(2)xU(1) ferromagnetic superconductivity described by the Lagrangian \cite{ap24,pla23}, 
\begin{gather}
\cL =-|{\cal D}_\mu \phi|^2 -\frac{\lambda}{2}\big(|\phi|^2
-\frac{\mu^2}{\lambda}\big)^2
-\frac14 F_\mn^2-\frac14 \G_\mn^2, \nn \\
\cD_\mu \phi =\big(\pd_\mu-i\frac{g}{2} A_\mu
-i\frac{g'}{2} \vsig \cdot \B_\mu \big) \phi
=(D_\mu -i\frac{g}{2} A_\mu) \phi, \nn\\
D_\mu \phi=(\pd_\mu
-i\frac{g'}{2} \vsig \cdot \B_\mu \big) \phi,
\label{2lag0}
\end{gather}
where $\phi=(\phi_{\uparrow},\phi_{\downarrow})$ 
is the spin doublet Cooper pair, $A_\mu$ and 
$\B_\mu$ are the ordinary electromagnetic U(1) gauge potential and the SU(2) magnon gauge potential, $F_\mn$ and $\G_\mn$ are 
the corresponding field strengths, $g$ and $g'$ are the coupling constants. Notice that the non-Abelian magnon interaction here which describes the spin flip interaction between $\phi_{\uparrow}$ and $\phi_{\downarrow}$ Cooper pairs is identical to the magnon interactionin (\ref{lag1}). The new thing is that here we have the U(1) electromagnetic interaction because 
the Cooper pair has charge $2e$, so that we have $g/2=2e$. 

The justification of this Lagrangian as 
the Lagrangian for the two-gap superconductors comes from the fact that, if we neglect the spin structure of the Cooper pair and view $\phi$ as ordinary spin singlet and neglect the spin-flip magnon interaction of the Cooper pair, 
the Lagrangian reduces to the ordinary Landau-Ginzburg Lagrangian. This tells that 
the Lagrangian can be viewed as a natural non-Abelian extention of the ordinary superconductor which contains the magnon gauge interaction. 

Expressing the Cooper pair doublet $\phi$ with 
the scalar Higgs field $\rho$ and the SU(2) unit doublet $\xi$ by
\begin{gather}
\phi = \frac{1}{\sqrt{2}} \rho~\xi,
~~~(\xi^\dag \xi = 1),
\label{xi}
\end{gather}
and using the Abelian decomposition (\ref{cdec})
we have
\begin{gather}
\cD_\mu \xi= \Big[-i\frac{g}{2} A_\mu 
-i\frac{g'}{2} (B_\mu' \n +\W_\mu) \cdot \vsig \Big]~\xi,  \nn\\
|\cD_\mu \xi|^2 =\frac{1}{8} (-gA_\mu+g'B_\mu')^2 
+\frac{g'^2}{4} \W_\mu^2.
\end{gather}
From this we can remove the SU(2) unit doublet 
$\xi$ completely from (\ref{2lag0}) and ``abelianize" it gauge independently \cite{prd80,prl81}
\begin{gather}
\cL = -\frac12 (\pd_\mu \rho)^2
-\frac{\lam}{8}\big(\rho^2-\rho_0^2 \big)^2 \nn\\
-\frac14 F_\mn^2 -\frac14 {G_\mn'}^2
-\frac12 \big|D_\mu' W_\nu -D_\nu' W_\mu \big|^2  \nn\\
-\frac{\rho^2}{8} \big((-gA_\mu+g'B_\mu')^2 
+2 g'^2 W_\mu^*W_\mu \big)  \nn\\
+i g' G_\mn' W_\mu^* W_\nu 
+ \frac{g'^2}{4}(W_\mu^* W_\nu -W_\nu^* W_\mu)^2,  \nn\\
D_\mu'=\pd_\mu +ig' B_\mu',
~~~W_\mu =\frac{1}{\sqrt 2} (W^1_\mu + i W^2_\mu).
\label{2lag1}
\end{gather}
This tells that the Lagrangian is made of the Abelian electromagnetic gauge fields $A_\mu$, 
the Abelian magnon gauge field $B_\mu'$, and 
a complex magnon gauge field $W_\mu$. To understand what happened to $\xi$, notice that 
the two Abelian gauge fields in the Lagrangian 
are not mass eigenstates. To express them in 
terms of mass eigenstates, we introduce 
the following photon-magnon mixing by
\begin{gather}
\left(\begin{array}{cc} \bA_\mu \\
Z_\mu  \end{array} \right)
=\frac{1}{\sqrt{g^2 +g'^2}} \left(\begin{array}{cc} 
g' & g \\ -g & g' \end{array} \right)
\left(\begin{array}{cc} A_\mu \\ B'_\mu
\end{array} \right)  \nn\\
= \left(\begin{array}{cc}
\cos \om & \sin \om \\
-\sin \om & \cos \om \end{array} \right)
\left(\begin{array}{cc} A_\mu \\ B_\mu'
\end{array} \right),
\label{mix}
\end{gather}
where $\om$ is the mixing angle. With this we can 
express the Lagrangian (\ref{2lag1}) by
\begin{gather}
\cL = -\frac12 (\pd_\mu \rho)^2
-\frac{\lam}{8}\big(\rho^2-\rho_0^2 \big)^2
-\frac14 {\bF_\mn}^2 -\frac14 Z_\mn^2 \nn\\
-\frac12 \big|(\bD_\mu +i \be\frac{g'}{g} Z_\mu)W_\nu 
-(\bD_\nu +i \be\frac{g'}{g} Z_\nu)W_\mu \big|^2  \nn\\
-\frac{\rho^2}{4} \big(g'^2 W_\mu^*W_\mu
+\frac{g^2+g'^2}{2} Z_\mu^2 \big)
+i \be (\bF_\mn +\frac{g'}{g}  Z_\mn) W_\mu^* W_\nu   \nn\\
+ \frac{g'^2}{4}(W_\mu^* W_\nu - W_\nu^* W_\mu)^2,
\label{2lag2}
\end{gather}
where 
\begin{gather}
\bF_\mn=\pd_\mu \bA_\nu-\pd_\nu \bA_\mu, 
~~~Z_\mn = \pd_\mu Z_\nu-\pd_\nu Z_\mu,  \nn\\
\bD_\mu=\pd_\mu+i \be \bA_\mu,   \nn\\
\be=\frac{gg'}{\sqrt{g^2+g'^2}}=g' \sin\om =g \cos\om.
\label{e}
\end{gather}	
This is the physical expression of the non-Abelian Landau-Ginzburg Lagrangian (\ref{2lag0}), which tells that the three degrees of $\xi$ are absorbed to $Z_\mu$ and $W_\mu$ to make them massive, 
so that the above non-Abelian Landau-Ginzburg theory is made of Higgs scalar $\rho$, massless and massive Abelian gauge bosons $\bA_\mu$ and $Z_\mu$, and massive complex $W_\mu$ magnon whose masses are given by 
\begin{gather}
M_H= {\sqrt \lam} \rho_0,   \nn\\
M_W=\frac{g'}{2} \rho_0,
~~~M_Z=\frac{\sqrt {g^2+g'^2}}{2} \rho_0.		
\end{gather} 
So, unlike the Abelian Landau-Ginzburg theory, 
it has three mass scales. 

The mixing (\ref{mix}) has deep implications. 
First, notice that the two Abelian gauge bosons
$\bA_\mu$ and $Z_\mu$ coming from the mixing could naturally be interpreted to represent the real photon 
and the neutral magnon. This immediately tells that 
the massive magnon $W_\mu$ carries the electric 
charge as well as the spin charge, so that it is 
doubly charged. This must be clear from (\ref{2lag2}), 
which shows that $W_\mu$ couples to both $\bA_\mu$ 
and $Z_\mu$. Obviously, $W_\mu$ acquires the electric charge from $\xi$ absorbing the charge carried by 
the doublet. This is unexpected. 

Now, we may ask which describes the real photon. Assuming that this two-gap superconductor has 
the massive photon which generates the well 
known Meissner effect, we may identify $Z_\mu$ 
as the massive photon. In this case the coupling constant of $Z_\mu$ which couples to $W_\mu$ in (\ref{2lag2}) should become $e$. From this we have (with $g=4e$) 
\begin{gather}
g'= \frac{\sqrt{1+\sqrt{65}}}{\sqrt 2}~e,
~~~\tan \om= \frac{4\sqrt 2}{\sqrt{1+\sqrt{65}}},
\end{gather} 
so that 
\begin{gather}
M_W =\frac{\sqrt{1+\sqrt{65}}}{2\sqrt 2} e \rho_0, \nn\\
M_Z =\frac{\sqrt{9+\sqrt{65}}}{2\sqrt 2} e \rho_0  \simeq 1.4 M_W.
\label{mass2}
\end{gather}
This tells that the penetration length of 
the off-diagonal magnon field is 1.4 times bigger than that of the photon. More importantly, 
the massless Abelian magnon generates a long 
range magnetic interaction in the non-Abelian superconductor, which could naturally explain 
the long range ferromagnetic order in the two-gap superconductor \cite{ap24,pla23}.

Notice that, when we switch off the three magnons
$A_\mu$ and $W_\mu$, the Lagrangian (\ref{2lag2}) reduces to 
\begin{gather}
\cL = -\frac12 (\pd_\mu \rho)^2
-\frac{\lam}{8}\big(\rho^2-\rho_0^2 \big)^2
-\frac14 Z_\mn^2 \nn\\
-\frac{g^2}{8} \rho^2 Z_\mu^2,
\label{lglag}
\end{gather}
which is the well known Abelian Landau-Ginzburg Lagrangian. This confirms that, in the limit when 
the magnon interaction becomes very weak, 
the two-gap ferromagnetic superconductor is able to descrive the ordinary superconductor. This assures that the two-gap ferromagnetic superconductivity can be viewed as an extention of the ordinary superconductivity \cite{ap24,pla23}.  

On the other hand, when we neglect the U(1) electromagnetic interaction in (\ref{2lag0}) and concentrate on the SU(2) magnon interaction, we can interprete the Lagrangian to describe the non-Abelian magnon interaction between spin up and down electron 
spin doublet $\phi$, with the massless Abelian magnon $\bA_\mu$ and massive non-Abelian magnon $\W_\mu$. 
And clearly this is formally identical to 
the non-Abelian magnon gauge interaction in 
the frustrated magnetic material discussed above. 
So, we may conclude that as far as the magnon gauge interaction is concerned, the two-gap ferromagnetic supperconductor and the frustrated magnetic material has the sam underlying physics. The main difference is that in the two-gap suparconductor the magnetic source field is given by the spin doublet, while in the frustrated magnetic material the magnetic source field is given by the spin triplet.

And here again the mass generation of the photon
$Z_\mu$ and magnon $\W_\mu$ takes place without 
any symmetry breaking. Moreover, the Lagrangian
(\ref{2lag0}) admits non-Abelian magnonic magnetic vortex and monopole, as well as the well known 
Abelian Abrikosov vortex \cite{ap24,pla23}.  

It should be mentioned that the non-Abelian 
gauge interaction similar to (\ref{2lag1}) has also been discussed in spin-orbit coupling in colinear magnets \cite{jin}, and in the transport of itinarant electrons in frustrated magnets 
and altermagnets before \cite {zar1,zar2}.   

The Lagrangian (\ref{2lag0}) was proposed to describe two-gap ferromagnetic superconductivity. But it should be mentioned that this Lagrangian could also describe the spin flip interaction of 
the electron, if we treat the electron as the charged spinon and identify the Higgs doublet $\phi$ as the charged spinon. In this case, however, we may have to interprete the massless gauge field $\bA_\mu$ as the real photon, so that here we have three massive magnons $Z_\mu$ and $\W_\mu$ which describe the the magnon gage interaction of the electron spin. This is a very interesting possibility worth to be pursued further in more detail. 

{\bf Three-gap (ferro)magnetic superconductor}---The above discussion strongly implies that the non-Abelian magnon gauge interaction could be an important interaction in condensed matter physics. To re-inforce this view, we now discuss the three-gap magnetic superconductor made of the charged spin triplet 
Cooper pairs which has the SU(2)xU(1) gauge 
interaction described by the following Lagrangian    
\begin{gather}
\cL' =-\frac12 |\cD_\mu \vPhi|^2 -\frac{\lambda}{2}\big(|\vPhi|^2
-\frac{\mu^2}{\lambda}\big)^2
-\frac14 F_\mn^2-\frac14 \G_\mn^2, \nn \\
\cD_\mu \vPhi =\big(D_\mu-i g A_\mu \big) \vPhi, \nn\\ 
D_\mu \vPhi=(\pd_\mu +g' \B_\mu \times \big) \vPhi,
\label{3lag0}
\end{gather}
where $\vPhi$ is the SU(2) spin triplet made of two electron spins which has the U(1) electromagnetic phase, 
\begin{gather}
\vPhi= \rho~e^{-i\theta}~\n,~~~\rho=|\vPhi|.
\label{Phi}
\end{gather}
Notice that just like (\ref{2lag0}), this Lagrangian reduces to the ordinary Landau-Ginzburg Lagrangia if we neglect the spin structure of 
the Cooper pair and remove the spin-flip interaction of the magnon. 

From (\ref{Phi}) we have 
\begin{gather}
\cD_\mu \vPhi =\big[\pd_\mu \rho -ig \rho (A_\mu +\frac1g \pd_\mu \theta) \big] e^{i\theta}~\n   \nn\\
+g' \rho~e^{i\theta}~\W_\mu \times \n,
\end{gather}
so that the lagrangian can be expressed by
\begin{gather}
\cL =- \frac12 (\pd_\mu \rho)^2 
-\frac{\lambda}{2}\big(\rho^2 -\rho_0^2 \big)^2
-\frac14 \bF_\mn^2 -\frac{g^2}{2} \rho^2 \bA_\mu^2   \nn\\
-\frac14 {G'}_\mn^2    
-\frac14 (\hD_\mu\W_\nu-\hD_\nu\W_\mu)^2
+g'^2 \rho^2 \W_\mu^2  \nn\\
-\frac{g}{2} G'_\mn \n \cdot (\W_\mu \times \W_\nu)
-\frac{g^2}{4} (\W_\mu \times \W_\nu)^2,   \nn \\
\bA_\mu =A_\mu +\frac1g \pd_\mu \theta. 
\label{3lag1}
\end{gather}
This tells that this Lagrangian again has the mass generation of the electromagnetic field $\bA_\mu$ and the off-diagonal mognon field $\W_\mu$, but the Abelian magnon $B'_\mu$ remains massless.  Moreover, unlike (\ref{2lag0}), here we have no photon-magnon mixing and $\W_\mu$ remains electrically neutral. Other than this, the non-Abelian 
magnon gauge interaction is almost identical to the above two cases. For example, here again just like the above two cases we have a massless magnon $B'_\mu$ which induces a long range magnetic interaction.  

Clearly this spin-triplet superconductor can also 
have similar interesting topological objects, 
the non-Abelian magnonic vortex and magnonic 
monopole as well as the Abelian Abrikosov vortex. 

{\bf Discussions}---In this paper we have discussed how the genuine non-Abelian gauge interactions can take place in condensed matters. We have discussed three possible categories of condensed matters, the frustrated magnetic material, two-gap ferromagnetic superconductors, and three-gap magnetic superconductors, where 
the non-Abelian magnonic gauge interaction could play an important role. But similar non-Aelian 
and non-Abelian gauge interaction could also play an important role in other condensed matters like the spin liquid. 

An important and inevitable consequence of these non-Abelian gauge interactions in condensed matters is the emergence of non-Abelian topological objects, in particular non-Abelian magnetic vortex and monopole. Indeed in all 
three cases discussed above, the non-Abelian topological objects play important roles. But 
in detail the topological objects are different. For example, the monopole in (\ref{lag1}) and (\ref{3lag0}) is the 'tHooft-Polyakov 
type \cite{thooft}. But the one in (\ref{2lag0}) is the Cho-Maison type, a hybrid between Dirac and 'tHooft-Polyakov monopoles \cite{plb97}. 
The existence of these topological objects strongly implies the existence of novel interactions between magnons and topological objects and the existence of the non-Abelian Meissner effect and in these condensed matters. 

We emphasize that the above magnon gauge interaction is, as far as we understand, the first spin-spin interaction mediated by the messenger bosons \cite{ap24,pla23}. So far the spin-spin interactions in physics have always been treated as an instantaneous action at a distance which 
has no messenger particle. This was strange, because in morden physics all fundamental interactions are mediated by messenger particles. Now our discussion tells that the spin-spin interaction can also be described as a normal interaction mediated by the messenger particles. 

In high energy physics the non-Abelian gauge interaction plays a central role, but in 
condensed matter physics this was not so. 
Now, the work by Zarzuela and Kim and our discussion in this paper could change this situation completely and make the non-Abelian gauge interaction a mainstream in condensed 
matter physics \cite{zarkim,ap24,pla23}. 
In this connection it should be pointed out that 
the Lagrangian (\ref{lag1}) is mathematically 
identical to what is known as the Georgi-Glashow 
Lagrangian, and (\ref{2lag0}) is identical to 
the Weinberg-Salam Lagrangian proposed to describe the electroweak interaction. But here the same Lagrangians could describe totally different physics in totally different surroundings. 
For example, the Higgs vacuum here is supposed 
to be of the order of meV, but in the standard model it becomes of 100 GeV, different by 
the factor $10^{14}$. This is remarkable. 

It should be emphasized that the non-Abelian magnonic gauge interactions discussed in this Letter is a theoretical proposition, so that 
they should be tested by experiments. We have provided enough theoretical motivation for 
the possible existence of such interactions in condensed matters, and experiments should test 
the viability of this theoretical proposition. 
This is an important task in experimental condensed matter physics. A first step might 
be to confirm the existence of the new 
non-Abelian Meissner effect in the above condensed matters by experiments. 
  
It is natural to expect similar non-Abelian 
gauge interactions could also emerge in atomic physics in two-component Bose-Einstein 
condensates \cite{pra05,and}. The details of 
the above discussions will be published in 
a separate paper \cite{cho}.

{\bf ACKNOWLEDGEMENT}

~~~This work is supported in part by the National Research Foundation of Korea funded by the Ministry of Science and Technology (Grant 2025-R1D1A1B0-7045163) and by Center for Quantum Spacetime, Sogang University, Korea.


\begin{references}
\bibitem{wen} See, e.g., P.A. Lee, N. Nagaosa, 
and X.G. Wen, Rev. Mod. Phys. {\bf 78}, 17 (2006), 
and refereces therin. 	
\bibitem{zhong} F. Zhong et al. Science {\bf 302}, 92 (2003).
\bibitem{prb05} Y.M. Cho, Phys. Rev. {\bf B72}, 212516 (2005).
\bibitem{prb06} Y.M. Cho and Pengming Zhang, 
Phys. Rev. {\bf B73}, 180506 (2006).

\bibitem{epjb08} Y.M. Cho and P.M. Zhang, Euro Phys. J. {\bf B65}, 155 (2008). 
\bibitem{ap24} Y. M. Cho and Franklin H. Cho, 
Ann. Phys. {\bf 460},169573 (2024).
\bibitem{pla23} Y. M. Cho and Franklin H. Cho, Phys. Lett. {\bf A472}, 128793 (2023).

\bibitem{zarkim} R. Zarzuela and S. K. Kim, 
Phys. Rev. Lett. {\bf 134}, 186701 (2025).

\bibitem{prd80} Y. M. Cho, Phys. Rev. {\bf D21}, 
1080 (1980). See also Y. S. Duan and M. L. Ge, 
Sci. Sinica {\bf 11},1072 (1979).
\bibitem{prl81} Y. M. Cho, Phys. Rev. Lett. {\bf 46}, 302 (1981); Phys. Rev. {\bf D23}, 2415 (1981).

\bibitem{fadd} L. Faddeev and A. Niemi, Phys. Rev. Lett. {\bf 82}, 1624 (1999); Phys. Lett. 
{\bf B449}, 214 (1999).
\bibitem{shab}S. Shabanov, Phys. Lett. {\bf B458}, 
322 (1999); {\bf B463}, 263 (1999); H. Gies, Phys. 
Rev. {\bf D63}, 125023 (2001).
\bibitem{kondo} K. Kondo, S. Kato, A. Shibata, 
and T. Shinohara, Phys. Rep. {\bf 579}, 1 (2015).

\bibitem{jin} P.Q. Jin, Y.Q. Li, and . Zhang,
J. Phys. {\bf A39}, 7115 (2006).
\bibitem{zar1} R. Zarzuela and J. Sinova, 
Phys. Rev. {\bf B105}, 024423 (2022).\bibitem{zar2} R. Zarzuela, R. Jaeschke-Ubiergo, O. Gomonay, L. Smejcal,  
and J. Sinova, Phys. Rev. {\bf B111}, 064422 (2025).

\bibitem{thooft} G. 'tHooft, Nucl. Phys. 
{\bf B79}, 276 (1974); A. Polyakov, JETP Lett. {\bf 20}, 194 (1974).
\bibitem{plb97} Y. M. Cho and D. Maison, Phys. Lett. {\bf B391}, 360 (1997).

\bibitem{pra05} Y.M. Cho, Hyojoong Kim, and 
Pengming Zhang, Phys. Rev. {\bf A72}, 063603 (2005).
\bibitem{and} S.N. Andrianov and S.A. Moiseev, 
Phys. Rev. {\bf A90}, 042303 (2014).

\bibitem{cho} Y.M. Cho and Franklin H. Cho, 
to be published.

\end{references}
\end{document}